\headline={\ifnum\pageno>1 \hss \number\pageno\ \hss \else\hfill \fi}
\pageno=1
\nopagenumbers
\centerline{\bf AN EXPLICIT CONSTRUCTION OF CASIMIR OPERATORS}
\centerline{\bf AND EIGENVALUES : II }
\vskip 15mm
\centerline{\bf H. R. Karadayi and M. Gungormez}
\centerline{Dept.Physics, Fac. Science, Tech.Univ.Istanbul }
\centerline{ 80626, Maslak, Istanbul, Turkey }
\centerline{ Internet: karadayi@sariyer.cc.itu.edu.tr}
\vskip 10mm
\centerline{\bf{Abstract}}
\vskip 10mm

It is given a way of computing Casimir eigenvalues for Weyl orbits as well as
for irreducible representations of Lie algebras. A $\kappa(s)$ number of
polinomials of rank N are obtained explicitly for $A_N$ Casimir operators of
order s where $\kappa(s)$ is the number of partitions of s into positive
integers except 1. It is also emphasized that these eigenvalue polinomials
prove useful in obtaining formulas to calculate weight multiplicities and
in explicit calculations of the whole cohomology ring of Classical and also
Exceptional Lie algebras.

\vskip 15mm
\vskip 15mm
\vskip 15mm
\vskip 15mm
\vskip 15mm
\vskip 15mm
\vskip 15mm
\vskip 15mm

\hfill\eject

\vskip 3mm
\noindent {\bf{I.\ INTRODUCTION}}
\vskip 3mm
In a previous paper {\bf [1]} which we refer (I) throughout the
work, we establish the {\bf most general} explicit forms of 4th and 5th order
Casimir operators of $A_N$ Lie algebras. By starting from this point,
we want to develop a framework which makes possible to calculate, for the
irreducible representations of $A_N$ Lie algebras, the eigenvalues of Casimir
operators in any order. Extensions are also possible to any other classical
or exceptional Lie algebra because any Lie algebra has always an appropriate
subalgebra of type $A_N$.

For a Casimir operator \ I(s) of degree \ s, the eigenvalues for a D-dimensional
representation are known to be calculated in the following form:
$$ {1 \over D} \ \ \ Trace(I(s)) \ \ .  \eqno(I.1) $$
A direct calculation of (I.1) could become problematic in practice as
the dimension of representation grows high. Additionally to the ones given
in (I), we give here some further works {\bf [2]} dealing with this problem.

A second essential problem arisen here is due to the fact that one must also
calculate weight multiplicities for representations comprising more than one
Weyl orbit. This latter problem is known to be solved by formulas which are
due to Kostant and Freudenthal {\bf[3]} and it is at the root of Weyl-Kac
character formulas {\bf [4]}. Although they are formally explicit, these two
formulas are of recursive character and hence they exhibit problems in
practical calculations. One could therefore prefers to obtain a {\bf functional
formula} in calculating weight multiplicities. This will be dealt in a
subsequent paper.

It is known, on the other hand, that {\bf trace operations} can be defined
{\bf [5]} in two equivalent ways one of which is nothing but the explicit
matrix trace. An expression like (I.1) could therefore not means for a Weyl
orbit, {\bf in general}. We instead want to extend the concept of Casimir
eigenvalue to Weyl orbits. As we have introduced in an earlier work
{\bf [6]}, we replace (I.1) with the following {\bf formal} definition:
$$ ch_s(\Pi) \equiv \sum_{\mu \in \Pi} (\mu)^s  \eqno(I.2) $$
where $\Pi$ is a Weyl orbit and the sum is over all weights $\mu$
included within $\Pi$. The powers of weights in (I.2) are to be thought
of as s-times products
$$  (\mu)^s=\overbrace{\mu\times \mu\times ...\times \mu}^{s~times} \ \ . $$
Note here that (I.2) is defined not only for Weyl orbits or
representations but it means also for any collection of weights.
We will mainly show in what follows how (I.2) gives us a way to obtain
eigenvalues of a Casimir operator. Due to a permutational lemma given in
section (II), the procedure works out especially for $A_N$ Lie algebras.
It will however be seen in a subsequent paper that it is generalized to any
Classical or Exceptional Lie algebra. In section (III), we will give a general
formula of calculating $ch_s(\Pi)$ by the aid of this permutational lemma.
An efficient way of using this formula is due to reduction rules which are
explained in section (IV) and the polinomials representing Casimir eigenvalues
will be given in section (V) and also in appendix.2. We will show in section
(VI) that the two formula (I.1) and (I.2) are in fact in coincidence.

\vskip 3mm
\noindent {\bf{II.\ A PERMUTATIONAL LEMMA FOR $A_N$ WEYL ORBITS}}
\vskip 3mm
In this section, we give, for $A_N$ Lie algebras, a permutational lemma
which says that, {\bf modulo permutations, there is one-to-one correspondence
between the Weyl chamber and the Tits cone} {\bf [7]}. As will be explained
below, such a correspondence appears only when one reformulates everything
in terms of the so-called {\bf fundamental weights}.

For an excellent study of Lie algebra technology we refer the book of
Humphreys {\bf [8]}. We give, however, some frequently used concepts here.
In describing the whole {\bf weight lattice} of a Lie algebra of rank \ N,
the known picture will be provided by {\bf simple roots $\alpha_i$} and
{\bf fundamental dominant weights $\lambda_i$} where indices like
$i_1,i_2,.. $ take values from the set {\bf $I_\circ \equiv \{1,2,.. N\}$}.
Any dominant weight $ \Lambda^+ $ can then be expressed by
$$ \Lambda^+ = \sum_{i=1}^{N} r_i \  \lambda_i \ \ , \ \ r_i \in Z^+  \eqno(II.1) $$
where $Z^+$ is the set of positive integers including zero.
We know that a Weyl orbit $\Pi$ is stable under the actions of Weyl group of
Lie algebra. This means that all weights within a Weyl orbit are equivalent
under the actions of Weyl group and they can be obtained from any one of them
by performing Weyl conjugations one-by-one. We thus obtain a description of
the whole weight lattice of which any weight is given by
$$ \mu = m_1 \ \lambda_1 + m_2 \ \lambda_2 + ... + m_N \ \lambda_N \ \ , \ \
\pm m_i \in Z^+ \ \ .  \eqno(II.2) $$

Our way of thinking of a Weyl orbit is, on the other hand, based on the fact
that {\bf Weyl reflections can be replaced by permutations} for $A_N$ Lie
algebras. It is seen in the following that essential figures for this are
{\bf fundamental weights} ${\bf \mu_I}$ which we introduced {\bf [9]}
some fifteen years ago:
$$ \eqalign{
\mu_1 &\equiv \lambda_1 \cr
\mu_i &\equiv \mu_{i-1} - \alpha_{i-1} \ \ , \ \ i = 2,3,.. N+1. } \eqno(II.3) $$
Indices like $I_1,I_2,...$ take values from the set
$S_\circ \equiv \{1,2,.. N,N+1\}$. Recall here that the weights defined in
(II.3) are nothing but the weights of (N+1)-dimensional fundamental
representations of $A_N$ Lie algebras. To prevent confusion, note here that
some authors prefer to call $\lambda_i$'s fundamental weights.Though there
are N+1 number of fundamental weights  $\mu_I$, they are not completely
linear independent due to the fact that their sum is zero. The main
observation is, however, that (II.2) replaces with
$$ \mu = q_1 \ \mu_{I_1} + q_2 \ \mu_{I_2} + .. + q_{N+1} \ \mu_{I_{N+1}} \eqno(II.4) $$
when one reformulates in terms of N+1 fundamental weights. The conditions
$$ I_1 \neq I_2 \neq .. \neq I_{N+1} \eqno(II.5) $$
must be taken into account for each particular weight (II.4) and one can
always assume that
$$ q_1 \geq q_2 \geq ... \geq q_{N+1} \geq 0 \ \ . \eqno(II.6) $$
(II.6) receives here further importance in the light of following lemma:

Let {\it P(N)} be the weight lattice of $A_N$ Lie algebra. A dominant
weight $\Lambda^+ \in {\it P(N)} $ has always the form of
$$ \Lambda^+ = q_1 \ \mu_1 + q_2 \ \mu_2 + .. + q_{N+1} \ \mu_{N+1} \eqno(II.7) $$
and hence the whole Weyl orbit $ \Pi(\Lambda^+) $ is obtained by
permutations of (II.7) over N+1 fundamental weights. In the basis of fundamental
weights all weights of the Weyl orbit $\Pi(\Lambda^+)$ are thus seen
in the common form (II.4) where all indices $I_k $ take values from the set
$S_\circ$ together with the conditions (II.5).

Although it is not in the scope of this work, demonstration of lemma
is a direct result of the definitions (II.3). It will be useful to realize the
lemma further in terms of (N+1)-tuples which re-define (II.7) in the form
$$ \Lambda^+ \equiv (q_1,q_2, .. q_{N+1}) \ \ . \eqno(II.8) $$
Then every elements $\mu \in \Pi(\Lambda^+)$ corresponds to a permutation of $q_i's$:
$$ \mu = (q_{I_1},q_{I_2}, .. ,q_{I_{N+1}}) $$.

To this end, let us choose a weight
$$ - \lambda_1 + 2~ \lambda_2 - \lambda_3 + \lambda_4 + \lambda_5 -
\lambda_6 + \lambda_7  \eqno(II.9) $$
which is expressed in the conventional form (II.2). By taking inverses
$$ \lambda_i \equiv \mu_1 + \mu_2 + ... + \mu_i \ \ ,
\ \ i \in I_\circ \eqno(II.10) $$
of (II.2), we can re-express (II.9) as
$$ 2 \mu_1 + 3 \ \mu_2 + \mu_3 + 2 \ \mu_4 + \mu_5 + \mu_7 \eqno(II.11) $$
which says us that
$$ - \lambda_1 + 2~ \lambda_2 - \lambda_3 + \lambda_4 + \lambda_5 - \lambda_6
+ \lambda_7 \in \Pi(\lambda_1 + \lambda_3 + \lambda_6) \ \ . $$
It is obvious that this last knowledge is not so transparent in (II.9).

One must further emphasize that the lemma allows us to know the dimensions
of Weyl orbits directly from their dominant representatives. For this and
further use, let us re-consider (II.1) in the form
$$ \Lambda^+ \equiv u_1 \ \lambda_{i_1} + u_2 \ \lambda_{i_2} + ... +
u_\sigma \ \lambda_{i_\sigma} \ \ , \ \ u_\sigma \in Z^+ - {0}   \eqno(II.12) $$
with
$$ i_1 \leq i_2 \leq ... \leq i_\sigma \ \ ,
\ \ \sigma = 1,2, .. N.     \ \ . \eqno(II.13) $$
Then, it is seen that the number of weights within a Weyl orbit
$ \Pi(\Lambda^+) $ is
$$ dim\Pi(\Lambda^+) = { (N+1)! \over \xi(\Lambda^+) \
(N+1-i_\sigma)! }   \eqno(II.14)  $$
where
$$ \xi(\Lambda^+) \equiv \prod_{j=1}^\sigma (i_j-i_{j-1})! \ \ ,
\ \ i_0 \equiv 0 \ \ . \eqno(II.15) $$
We therefore assume in the following that $dim\Pi(\Lambda)$ is always known
to be a polinomial of rank \ N.

\vskip 3mm
\noindent {\bf{III.\ EIGENVALUES FOR WEYL ORBITS}}
\vskip 3mm
As is mentioned above, eigenvalues are, in fact, known to be defined for
representations. A representation $ R(\Lambda^+) $ is, on the other hand,
determined from its {\bf orbital decomposition}:
$$ R(\Lambda^+) = \Pi(\Lambda^+) \ \ + \sum_{\lambda^+ \in Sub(\Lambda^+)}
m(\lambda^+ < \Lambda^+) \ \ \Pi(\lambda^+) \eqno(III.1) $$
where $Sub(\Lambda^+)$ is the set of all sub-dominant weights of $\Lambda^+$
and $m(\lambda^+ < \Lambda^+)$'s are multiplicities of weights
$\lambda^+$ within the representation $ R(\Lambda^+) $. Once a convenient
definition of eigenvalues is assigned to $ \Pi(\lambda^+) $ for
$\lambda^+ \in Sub(\Lambda^+)$, it is clear that this also means for the
whole $R(\Lambda^+)$ via (III.1). In the rest of this section, we then
show how definition (I.2) can be used to obtain {\bf orbit eigenvalues} as
N-dependent polinomials.

Let us now make some definitions which are used frequently for description
of {\bf symmetric polinomials} encountered in the root expansions which take
place heavily in the recently studied electromagnetically dual supersymmetric
theories {\bf [10]}. These will, of course, be given here
in terms of fundamental weights \ $ \mu_I $. The essential role will be played
by {\bf generators}
$$ \mu(s) \equiv \sum_{I=1}^{N+1} (\mu_I)^s \ \ , \ \ s = 1,2, ... \eqno(III.2) $$
and their reductive generalizations
$$ \mu(s_1,s_2, .. ,s_k) \equiv \sum_{I_1,I_2, .. I_k=1}^{N+1} (\mu_{I_1})^{s_1} (\mu_{I_2})^{s_2} ...\
(\mu_{I_k})^{s_k} \ \ . \eqno(III.3) $$
For (III.3), the conditions
$$ s_1 \geq s_2 \geq ... \geq s_k \eqno(III.4) $$
are always assumed and no two of indices $ I_1,I_2, .. I_k $
shall take the same value for each particular monomial. Note also that
$ \mu(s,0,0, .. 0) = \mu(s) . $

As the first step, we now make the suggestion, in view of (I.2), that
orbit eigenvalues can be conveniently calculated by decomposing $ ch_s(\Pi) $
in terms of quantities defined in (III.3) and this provides us the possibility
to calculate orbit eigenvalues with the same ability regardless

(  i) the rank N of algebra,

( ii) the dimension $ dimR(\Lambda^+,N) $ of irreducible representation,

(iii) the order s of Casimir element.

\noindent To give our results below, we will assume that the set
$$ s/k \equiv \{ s_1,s_2, ... ,s_k \} \eqno(III.5)  $$
represents, via (III.4), all partitions
$$ s = s_1 + s_2 + ... + s_k \ \ \ , \ \ \ s \geq k  $$
of positive integer s to k-number of positive integers $s_1,s_2, .. s_k$.
It is useful to remark here that each particular partition participating
within a \ s/k \ gives us, modulo (N+1), a dominant weight in P(N) and the 
whole subdominant chain $Sub(s \ \lambda_1)$ is in one-to-one correspondence 
with the partitions within a s/k. This must always be kept in mind in the 
following considerations.

On the other hand, instead of (II.1), it is crucial here to use (II.7) in the
form
$$ \Lambda^+ \equiv \sum_{i=1}^\sigma \ q_i \ \mu_i \eqno(III.6) $$
where $\sigma = 1,2, .. N+1$. Note here that this is another form of (II.12).
Due to permutational lemma given above, we now know that all weights of a
Weyl orbit are specified with the same parameters $q_i, (i=1,2, .. \sigma)$.
It is only of this fact which allows us to obtain the following formula in
expressing orbital eigenvalues:
$$ \Omega_s(q_1,q_2, ... ,q_\sigma,N) ={1 \over (N+1-\sigma)!} \ \
\sum_{k=1}^\sigma (N+1-k)! \ \xi(s/k) \ Factors(s/k) \eqno(III.7) $$
where we define, for all possible partitions (s/k),
$$ Factors(s/k) \equiv M(s_1,s_2, ... ,s_k) \ q(s_1,s_2, ... ,s_k) \
\mu(s_1,s_2, ... ,s_k) \eqno(III.8)  $$
and the multinomial
$$ M(s_1,s_2, ... ,s_k) \equiv {(s_1 + s_2 + ... + s_k)! \over s_1! s_2! ... s_k!} \ \  $$
together with the condition that
$$ M(s_1,s_2, ... ,s_k) \equiv 0 \ \ \ for \ \ \ s < k . \eqno(III.9) $$
$\xi(s/k)$ here is defined as in (II.15) because, as we remark just above,
any permutation within a s/k determines a dominant weight.
As in exactly the same way in (III.3), we also define
$$ q(s_1,s_2, .. ,s_k) \equiv \sum_{s_1,s_2, .. s_k=1}^\sigma (q_{I_1})^{s_1}
(q_{I_2})^{s_2} ...\ (q_{I_k})^{s_k} \ \ . \eqno(III.10) . $$

After all, one obtains a direct way to compute (I.2) in the form
$$ ch_s(\Lambda^+,N) = {1 \over \xi(\Lambda^+)} \
\Omega_s(q_1,q_2, ... ,q_k,N)  \eqno(III.11) $$
for all $q_1 \geq q_2 \geq .. \geq q_k$. For cases which we consider in this
work, we will give in appendix.1 some exemplary expressions extracted
from (III.7).

\hfill\eject

\vskip3mm
\noindent{\bf {IV. \ REDUCTION FORMULAS}}
\vskip3mm
Although it has an explicit form, the simplicity of formula (III.7) is not so
transparent to an experienced eye looking for its advanced applications. This
point can be recovered by recursively reducing the quantities (III.9) up to
generators $\mu(s)$ defined in (III.2). We call these {\bf reduction rules}.
We will only give the ones which we need in the sequel. It would however be
useful to mention about some of their general features. As is known,
elementary Schur functions ${\bf S_k(x)}$ are defined by expansions
$$ \sum_{k \in Z^+} \ S_k(x) \  z^k \equiv exp\sum_{k=1}^\infty \ x_k \ z^k \eqno(IV.1) $$
with the following explicit expressions:
$$ S_k(x) = \sum_{k_1+2 \ k_2+3 \ k_3 .. =k} \ {x_1^{k_1} \over k_1!} \ {x_2^{k_2} \over k_2!} ... \ \ \ , \ \ \ k>0 \ . \eqno(IV.2) $$
The complete symmetric functions $h_k(\mu_1,\mu_2, .. \mu_N)$ are defined,
on the other hand, by
$$ \prod_{i=1}^N \ {1 \over (1-z \ \mu_i)} \equiv \sum_{k \geq 0}
h_k(\mu_1,\mu_2, .. \mu_N) \ z^k \ \ . \eqno(IV.3)  $$
It can be easily shown that the known equivalence
$$ h_k(\mu_1,\mu_2, .. \mu_N) \equiv S_k(x) \eqno(IV.4)  $$
is now conserved by the reduction rules with the aid of a simple replacement
$$ \mu(s) \rightarrow s \ x_s \ \ . $$
A simple but instructive example concerning (IV.4) for k=4 is
$$ h_4(\mu_1,\mu_2,\mu_3,\mu_4) = \mu(4) + \mu(3,1) + \mu(2,2) + \mu(2,1,1) +
\mu(1,1,1,1) \eqno(IV.5) $$
with the corresponding reduction rules
$$ \eqalign{q&(1,1,1,1)={1 \over 24} \ q(1)^4 - {1 \over 4} \ q(1)^2 \ q(2) +
{1 \over 8} \ q(2)^2 + {1 \over 3} \ q(1) \ q(3) - {1 \over 4} \ q(4) \ \ , \cr
q&(2,1,1)={ 1\over 2} \ q(1)^2 q(2) - { 1 \over 2} \ q(2)^2 - q(1) \ q(3) + q(4) \ \ , \cr
q&(3,1)=q(1) \ q(3) - q(4)  \ \ , \cr
q&(2,2)={ 1\over 2} \ q(2)^2 - { 1\over 2} \ q(4) \ \ . }  \eqno(IV.6) $$
For other cases of interest, the reduction rules will be given in appendix.1
respectively for the partitions of 5,6 and 7.

\vskip3mm
\noindent{\bf {V.\ EXISTENCE OF EIGENVALUE POLINOMIALS}}
\vskip3mm
After all these preparations, we are now in the position to bring out the most
unexpected part of work. This is the possibility to extend (III.11) directly
for irreducible representations as well as Weyl orbits. We will show in a
subsequent work that this gives us the possibility to obtain infinitely many
functional formulas to calculate weight multiplicities and also to make
explicit calculations of nonlinear cohomology relations which are known to be
exist {\bf [11]} for classical and exceptional Lie algebras.

In view of the fact that $\mu(1) \equiv 0$, one can formally decompose (III.11)
in the form
$$ ch_s(\Lambda^+,N) \equiv \sum_{s/k} cof_{s_1s_2 .. s_k}(\Lambda^+,N) \
\mu(s_1) \mu(s_2) .. \mu(s_k)  \eqno(V.1) $$
and this allows us to define a number of polinomials
$$ P_{s_1 s_2 .. s_k}(\Lambda^+,N) \equiv {cof_{s_1 s_2 .. s_k}(\Lambda^+,N) \over
cof_{s_1 s_2 .. s_k}(\lambda_k,N)} \ {dimR(\lambda_k,N) \over dimR(\Lambda^+,N)} \
P_{s_1 s_2 .. s_k}(\lambda_k,N) \ \ . \eqno(V.2) $$
Note here that
$$ cof_{s_1 s_2 .. s_k}(\lambda_i,N) \equiv 0 \ \ \ , \ \ \ i < k  \eqno(V.3) $$
and also
$$ dimR(\lambda_i,N) = M(N+1,i) \ \ \ , \ \ \ i=1,2, .. N. \ \ . \eqno(V.4)  $$

To proceed further, we will work on the explicit example of 4th order for
which (V.1) and (V.2) give
$$ ch_4(\Lambda^+,N) \equiv cof_4(\Lambda^+,N) \ \mu(4) +
                      cof_{22}(\Lambda^+,N) \ \mu(2)^2 \ \ , \eqno(V.5)  $$

$$ P_4(\Lambda^+,N) \equiv {cof_4(\Lambda^+,N) \over cof_4(\lambda_1,N)} \
{dimR(\lambda_1,N) \over dimR(\Lambda^+,N)} \ P_4(\lambda_1,N) \ \ , \eqno(V.6) $$
and
$$ P_{22}(\Lambda^+,N) \equiv {cof_{22}(\Lambda^+,N) \over cof_{22}(\lambda_2,N)} \
{dimR(\lambda_2,N) \over dimR(\Lambda^+,N)} \ P_{22}(\lambda_2,N)  \eqno(V.7) $$
N dependences are explicitly written above. The main observation here is to
change the variables $r_i$ of (II.1):
$$ 1 \ + \ r_i \equiv \theta_i \ - \ \theta_{i+1} \ \  \eqno(V.8) $$
and to suggest the decompositions
$$ \eqalign{
P_4(\Lambda^+,N) = &k_4(1,N) \ \Theta(4,\Lambda^+,N)   \ \ + \cr
                   &k_4(2,N) \ \Theta(2,\Lambda^+,N)^2 \ + \cr
                   &k_4(3,N) \ \Theta(3,\Lambda^+,N)   \ \ + \cr
                   &k_4(4,N) \ \Theta(2,\Lambda^+,N)   \ \ + \cr
                   &k_4(5,N) }                           \eqno(V.9) $$
and
$$ \eqalign{
P_{22}(\Lambda^+,N) = &k_{22}(1,N) \ \Theta(4,\Lambda^+,N)   \ \ + \cr
                      &k_{22}(2,N) \ \Theta(2,\Lambda^+,N)^2 \ + \cr
                      &k_{22}(3,N) \ \Theta(3,\Lambda^+,N)   \ \ + \cr
                      &k_{22}(4,N) \ \Theta(2,\Lambda^+,N)   \ \ + \cr
                      &k_{22}(5,N) }        \ \ \ .            \eqno(V.10) $$
As in (III.2) or (III.10), we also define here the generators
$$ \Theta(s,\Lambda^+,N) \equiv \sum_{i=1}^{N+1} \ (\theta_i)^s  \ \ \ . \eqno(V.11) $$
It is seen then that (V.9) and (V.10) are the most general forms compatible
with $\Theta(1,\Lambda^+,N) \equiv 0$. What is significant here is the
possibility to solve equations (V.6) and (V.7) in view of assumptions (V.9)
and (V.10) but with coefficients $k_4(\alpha,N),k_{22}(\alpha,N)$ {\bf which
are independent of $\Lambda^+$} for $ \alpha = 1, .. ,5 $. By examining for a
few simple representations, one can easily obtain the following {\bf non-zero}
solutions for these coefficients:
$$ \eqalign{ &k_4(1,N)={720 \over g_4(N)} \ (N^2 + 2 N + 2) \ k_4(5,N) \cr
             &k_4(2,N)=- \ {720 \over g_4(N) \ (N + 1)} \
                                (2 \ N^2 + 4 \ N - 1) \ k_4(5,N) } \eqno(V.14) $$
and
$$ \eqalign{
&k_{22}(1,N)=- \ {1440 \over g_{22}(N)} \ (2 \ N^2 + 4 \ N - 1) \ k_{22}(5,N) \cr
&k_{22}(2,N)={720 \over g_{22}(N) \ (N + 1)} \ (N^4 + 4 \ N^3 - 8 \ N + 13) \ k_{22}(5,N) \cr
&k_{22}(4,N)=- \ {120 \over g_{22}(N)} \ (N - 2) \ (N - 1) \ (N + 1)^2 \ (N + 3) \ (N + 4) \ k_{22}(5,N) } \eqno(V.15) $$
where
$$ \eqalign{
&g_4(N) \equiv \prod_{i=-2}^4 (N+i) \cr
&g_{22}(N) \equiv g_4(N) \ (5 \ N^2 + 10 \ N + 11) \ \ . } \eqno(V.16)  $$

The calculations goes just in the same way for orders 5,6 and 7 and hence
we directly give our solutions in appendix.2.

\vskip3mm
\noindent{\bf {VI.\ CONCLUSIONS}}
\vskip3mm
In (I), we have obtained the most general formal operators representing
4th and also 5th order Casimir invariants of $A_N$ Lie algebras. By comparing
with the ones appearing in litterature, they are the most general in the sense
that both are to be expressed in terms of two free parameters. As is shown
in (I), all coefficient polinomials of 4th order Casimir operators are expressed
in terms of u(1) and u(2) while those of 5th order Casimirs are v(1) and v(2).
As is also emphasized there, the existence of two free parameters for both
cases can be thought of as related with the partitions 4=2+2 and 5=3+2.
Recall here the polinomials $P_4$ and $P_{22}$. This gives us the possibility
to calculate the trace forms (I.1) directly in any matrix representation of
$A_N$ Lie algebras. These trace calculations are straigtforward and show that
eigenvalues of 4th order Casimir operators have the form of an explicit
polinomial which depends on the rank N and two free parameters u(1) and u(2).
It is thus seen that there are always appropriate choices of parameters
u(1) and u(2) in such a way that this same polinomial reproduces
$P_4(\Lambda^+,N)$ or $P_{22}(\Lambda^+,N)$ as given in (V.9) and (V.10).
The same is also true for 5th order Casimirs. With the appropriate choice
$$ k_4(5,N) \equiv {1 \over 6!} \ (N + 1)^2 \ (N + 2) \ (N + 3) \ (N + 4) \eqno(VI.1)  $$
in (V.9) it is sufficient to take
$$ u(1) = 1 \ \ \ , \ \ \ u(2) = {3 \ N-8 \over 3 \ N} \eqno(VI.2) $$
in order to reproduce
$$ {1 \over {_D}} \ Trace(I(4)) \equiv P_4(\Lambda^+,N) $$
with $dimR(\Lambda^+,N) = D$. The data for other cases of interest are
$$ \eqalign{
&k_{22}(5,N) \equiv {1 \over 6!} \ (5 \ N^2 + 10 \ N + 11) \ (N + 1) \ (N + 2) \ (N + 3) \ (N + 4) \cr
&u(1) = 1 \ \ \ , \ \ \ u(2) = {2 \over 3} \ {2\ N^2 + N + 2 \over N \ (N + 1)}  }  \eqno(VI.3)     $$
for
$$ {1 \over {_D}} \ Trace(I(4)) \equiv P_{22}(\Lambda^+,N) \ \ ,  $$
and
$$ \eqalign{
&k_5(2,N) \equiv -5 \ {(N + 1) \ (N^2 + 2 \ N - 1) \over N
\ (N - 1) \ (N - 2) \ (N - 3)} \cr
&v(1) = 1 \ \ \ , \ \ \ v(2) = {2 \ N - 5 \over 2 \ N}  }  \eqno(VI.4)     $$
for
$$ {1 \over {_D}} \ Trace(I(5)) \equiv P_{5}(\Lambda^+,N) \ \ ,  $$
and
$$ \eqalign{
&k_{32}(5,N) \equiv - {1 \over 12} \ {(N + 1)^3 \ (N + 4) \ (N + 5) \over N \ (N - 1)} \cr
&v(1) = 1 \ \ \ , \ \ \ v(2) = {(11 \ N + 5) \ (N - 1) \over 10 \ N \ (N + 1)}  }  \eqno(VI.5)     $$
for
$$ {1 \over {_D}} \ Trace(I(5)) \equiv P_{32}(\Lambda^+,N) \ \ .  $$

Now it is clear that, this would be a {\bf direct evidence} for equivalence
between the formal expressions (I.1) and (I.2). In result, it is seen that
one can obtain $\kappa(s)$ number of different polinomials
$P_{s_1,s_2, .. s_k}(\Lambda^+,N)$ representing eigenvalues of $A_N$ Casimir
operators I(s) of order s, with $\kappa(s)$ is the number of partitions of s
to all positive integers except 1. As is known from (I), this is just the
number of free parameters to describe the most general form of I(s).

\vskip3mm
\noindent{\bf {REFERENCES}}
\vskip3mm
\noindent [1] Karadayi H.R and Gungormez M: Explicit Construction of Casimir
Operators and Eigenvalues:I , submitted to J.Math.Phys.
\vskip 1mm
\noindent [2]   Braden H.W ; Jour.Math.Phys. 29 (1988) 727-741 and 2494-2498

\noindent Green H.S and Jarvis P.D ; Jour.Math.Phys. 24 (1983) 1681-1687
\vskip 1mm
\noindent [3]   Kostant B. ; Trans.Am.Math.Soc. 93 (1959) 53-73

\noindent Freudenthal H. ; Indag.Math. 16 (1954) 369-376 and 487-491

\noindent Freudenthal H. ; Indag.Math. 18 (1956) 511-514
\vskip 1mm
\noindent [4]   Kac.V.G ; Infinite Dimensional Lie Algebras, 3rd edition,
Cambridge University Press
\vskip 1mm
\noindent [5] the paragraph (3.29) in Carter, R.W: Simple Groups of
Lie Type, J.Wiley and sons (1972) N.Y
\vskip 1mm
\noindent [6]   Karadayi H.R, Jour.Math.Phys. 25 (1984) 141-144
\vskip 1mm
\noindent [7] the section 3.12 in ref.4
\vskip 1mm
\noindent [8] Humphreys J.E: Introduction to Lie Algebras and Representation
Theory , Springer-Verlag (1972) N.Y.
\vskip 1mm
\noindent [9] Karadayi H.R: Anatomy of Grand Unifying Groups ,
ICTP preprints (unpublished), IC/81/213 and 224
\vskip 1mm
\noindent [10] Kutasov D, Schwimmer A and Seiberg N: Chiral rings, Singularity
Theory and Electric-Magnetic Duality , {\bf hep-th/9510222}
\vskip1mm
\noindent [11] \ Borel A and Chevalley C: Mem.Am.Math.Soc. 14 (1955) 1

\noindent Chih-Ta Yen: Sur Les Polynomes de Poincare des Groupes de Lie
Exceptionnels, Comptes Rendue Acad.Sci. Paris (1949) 628-630

\noindent Chevalley C: The Betti Numbers of the Exceptional Simple Lie Groups,
Proceedings of the International Congress of Mathematicians, 2 (1952) 21-24

\noindent Borel A: Ann.Math. 57 (1953) 115-207

\noindent Coleman A.J: Can.J.Math 10 (1958) 349-356

\hfill\eject

\vskip3mm
\noindent{\bf {APPENDIX.\ 1}}
\vskip3mm
In this work, we consider the calculation of eigenvalues for $A_N$ Casimir
operators of orders s=4,5,6,7. It is however apparent that all our results
are to be accomplished as in exactly the same way and with the same
ability for all orders. The following applications of the formula (III.7)
will be instructive for all other cases of interest:
$$ \eqalign{
\Omega_4&(q_1,N) ={1 \over (N+1-1)!} \ \ (          \cr
& 1!   \ (N+1-1)! \ M(4) \ q(4) \ \mu(4) \ \ ) \ \ , }   \eqno(A1.1) $$
$$ \eqalign{
\Omega_4&(q_1,q_2,N) ={1 \over (N+1-2)!} \ \ (          \cr
& 1!   \ (N+1-1)! \ M(4) \ q(4) \ \mu(4) \ \ \  \ \ \ \ \ \ \ +       \cr
& 1!   \ (N+1-2)! \ M(3,1) \ q(3,1) \ \mu(3,1)  \     +       \cr
& 2!   \ (N+1-2)! \ M(2,2) \ q(2,2) \ \mu(2,2) \ \ \ ) \ \ , } \eqno(A1.2) $$
$$ \eqalign{
\Omega_4&(q_1,q_2,q_3,N) ={1 \over (N+1-3)!} \ \ (          \cr
& 1!   \ (N+1-1)! \ M(4) \ q(4) \ \mu(4) \ \ \  \ \ \ \ \ \ \ \ \ \ \ \ \ +       \cr
& 1!   \ (N+1-2)! \ M(3,1) \ q(3,1) \ \mu(3,1)  \ \ \ \ \ \ \     +       \cr
& 2!   \ (N+1-2)! \ M(2,2) \ q(2,2) \ \mu(2,2)  \ \ \ \ \ \ \     +       \cr
& 2!   \ (N+1-3)! \ M(2,1,1) \ q(2,1,1) \ \mu(2,1,1) \ \ ) \ \ , } \eqno(A1.3) $$
and for $k \geq 4$
$$ \eqalign{
\Omega_4&(q_1,q_2, .. ,q_\sigma,N) ={1 \over (N+1-\sigma)!} \ \ (        \cr
& 1!   \ (N+1-1)! \ M(4) \ q(4) \ \mu(4) \ \ \ \ \ \ \ \ \ \ \ \ \ \ \ \ \ \ +       \cr
& 1!   \ (N+1-2)! \ M(3,1) \ q(3,1) \ \mu(3,1) \ \ \ \ \ \ \ \ \ \     +       \cr
& 2!   \ (N+1-2)! \ M(2,2) \ q(2,2) \ \mu(2,2) \ \ \ \ \ \ \ \ \ \     +       \cr
& 2!   \ (N+1-3)! \ M(2,1,1) \ q(2,1,1) \ \mu(2,1,1) \ \        +       \cr
& 4!   \ (N+1-4)! \ M(1,1,1,1) \ q(1,1,1,1) \ \mu(1,1,1,1) ) \ \ . } \eqno(A1.4) $$

On the other hand, for an effective application of (III.7), it is clear that
one needs to reduce the generators $q(s_1,s_2, .. s_k)$ in terms of
$q(s)$'s. Following ones are sufficient within the scope of this work.
Together with the condition that $\mu(1) \equiv 0$, the similar ones are valid
also
for $\mu(s_1,s_2, .. s_k)$'s:
$$ \eqalign{q&(4,1)=q(1) \ q(4) - q(5) \cr
q&(3,2)=q(2) \ q(3) - q(5) \cr
q&(3,1,1)={1 \over 2} \ (q(1)^2 \ q(3) -  q(2) \ q(3) -  2 q(1) \ q(4) + 2 q(5) )    \cr
q&(2,2,1)={1 \over 2} \ (q(1) \ q(2)^2 - 2 \ q(2) \ q(3) -  q(1) \ q(4) + 2 \ q(5)  ) \cr
q&(2,1,1,1)={1 \over 6} \ (q(1)^3 \ q(2) -  3 \ q(1) \ q(2)^2 - 3 \ q(1)^2 \ q(3) + \cr
& \ \ \ \ \ \ \ \ \ \ \ \ \ \ \ \ \ \ \ \ \ \ \ \ \ \ \ \ \ \ \ \ \ \
5 \ q(2) \ q(3) + 6 \ q(1) \ q(4) -  6 \ q(5)  ) \cr
q&(1,1,1,1,1)={1 \over 120} \ (q(1)^5 - 10 \ q(1)^3 \ q(2) + 15 \ q(1) \ q(2)^2 +
20 \ q(1)^2 \ q(3) -  \cr
& \ \ \ \ \ \ \ \ \ \ \ \ \ \ \ \ \ \ \ \ \ \ \ \ \ \ \ \ \ \ \ \ \ \ \ \ \
20 \ q(2) \ q(3) -30 \ q(1) \ q(4) + 24 \ q(5) ) }  \eqno(A1.5) $$

Beyond order 5, we will give the rules recursively as in the following:
$$\eqalign{
&q(i_1,i_2) = q(i_1) \ q(i_2)\ - \ q(i_1 \ + \ i_2) \ \ \ \ \ \ i_1 \ > \ i_2 \cr
&q(i_1,i_1) = {1 \over 2} (q(i_1)^2 \ - \ q(i_1 \ + \ i_1) ) \cr
&q(i_1,i_2,i_2) = q(i_1) \ q(i_2,i_2)\ - \ q(i_1 \ + \ i_2,i_2) \ \ \ \ \ \ i_1 \ > \ i_2 \cr
&q(i_1,i_1,i_2) = {1 \over 2} (q(i_1) \ q(i_1,i_2)\ - \ q(i_1 \ + \ i_1,i_2) \ - \ q(i_1 \ + \ i_2,i_1) ) \ \ \ \ \ i_1 \ > \ i_2 \cr
&q(i_1,i_2,i_3) = q(i_1) \ q(i_2,i_3)\ - \ q(i_1 \ + \ i_2,i_3) \ - \ q(i_1 \ + \ i_3,i_2)\ \ \ \ \ i_1 \ > \ i_2 \ > i_3 \cr
&q(i_1,i_1,i_1) = {1 \over 3} (q(i_1) \ q(i_1,i_1)\ - \ q(i_1 \ + \ i_1,i_1) ) \cr
&q(i_1,i_2,i_2,i_2) = q(i_1) \ q(i_2,i_2,i_2)\ - \ q(i_1 \ + \ i_2,i_2,i_2) \ \ \ \ \ \ i_1 \ > \ i_2 \cr
&q(i_1,i_1,i_1,i_2) = {1 \over 3} (q(i_1) \ q(i_1,i_1,i_2)\ - \ q(i_1 \ + \ i_2,i_1,i_1) ) \ \ \ \ \ \ i_1 \ > \ i_2 \cr
&q(i_1,i_1,i_2,i_2) = {1 \over 2} (q(i_1) \ q(i_1,i_2,i_2)\ - \ q(i_1 \ + \ i_2,i_1,i_2) ) \ \ \ \ \ \ i_1 \ > \ i_2 \cr
&q(i_1,i_2,i_3,i_3) = q(i_1) \ q(i_2,i_3,i_3)\ - \ q(i_1 \ + \ i_2,i_3,i_3) \ - \ q(i_1 \ + \ i_3,i_2,i_3)\ \ \ \ \ i_1 \ > \ i_2 \ > i_3 \cr
&q(i_1,i_1,i_1,i_1) = {1 \over 4} (q(i_1) \ q(i_1,i_1,i_1)\ - \ q(i_1 \ + \ i_1,i_1,i_1) )  \cr
&q(i_1,i_2,i_2,i_2,i_2) = q(i_1) \ q(i_2,i_2,i_2,i_2)\ - \ q(i_1 \ + \ i_2,i_2,i_2,i_2) \ \ \ \ \ \ i_1 \ > \ i_2 \cr
&q(i_1,i_1,i_2,i_2,i_2) = {1 \over 2} (q(i_1) \ q(i_1,i_2,i_2,i_2)\ - \ q(i_1 \ + \ i_2,i_1,i_2,i_2) ) \ \ \ \ \ \ i_1 \ > \ i_2 \cr
&q(i_1,i_1,i_1,i_1,i_1) = {1 \over 5} (q(i_1) \ q(i_1,i_1,i_1,i_1)\ - \ q(i_1 \ + \ i_1,i_1,i_1,i_1) )  \cr
&q(i_1,i_2,i_2,i_2,i_2,i_2) = q(i_1) \ q(i_2,i_2,i_2,i_2,i_2)\ - \ q(i_1 \ + \ i_2,i_2,i_2,i_2,i_2) \ \ \ \ \ \ i_1 \ > \ i_2 \cr
&q(i_1,i_1,i_1,i_1,i_1,i_1) = {1 \over 6} (q(i_1) \ q(i_1,i_1,i_1,i_1,i_1)\ - \ q(i_1 \ + \ i_1,i_1,i_1,i_1,i_1) )  \cr
&q(i_1,i_1,i_1,i_1,i_1,i_1,i_1) = {1 \over 7} (q(i_1) \ q(i_1,i_1,i_1,i_1,i_1,i_1)\ - \ q(i_1 \ + \ i_1,i_1,i_1,i_1,i_1,i_1) )  } $$

\vskip3mm
\noindent{\bf {APPENDIX.\ 2}}
\vskip3mm
In section.V we show the way of extracting two eigenvalue polinomials which
are shown to be valid in 4th order. We repeat here the analysis in orders
s=5,6,7 and we give our solutions respectively for

(1) the 4 eigenvalue polinomials in order 7 \ (=5+2=4+3=3+2+2)

$$ \eqalign{
&k_7(1,N)=   {k_7(4,N) \over 14 \ (2 \ N^2 + 4 \ N - 15) } \ (N^4 + 4 \ N^3 + 41 \ N^2 + 74 \ N + 120)  \cr
&k_7(2,N)= - {k_7(4,N) \over 2 \ (N + 1) \ (2 \ N^2 + 4 \ N - 15) } \ (N^4 + 4 \ N^3 + 17 \ N^2 + 26 \ N - 96) \cr
&k_7(3,N)= - {k_7(4,N) \over 2 \ (N + 1) \ (2 \ N^2 + 4 \ N - 15) } \ (N^4 + 4 \ N^3 + 5 \ N^2 + 2 \ N + 60) } $$
\ \ \ \ \ \ \ \ \ \ \ \ \ \ \ \ \
\ \ \ \ \ \ \ \ \ \ \ \ \ \ \ \ \
$$ \eqalign{ P_7(\Lambda^+,N) &=
   \ k_7(1,N) \ \Theta(7,\Lambda^+,N)                           \cr
&+ \ k_7(2,N) \ \Theta(5,\Lambda^+,N) \ \Theta(2,\Lambda^+,N)   \cr
&+ \ k_7(3,N) \ \Theta(4,\Lambda^+,N) \ \Theta(3,\Lambda^+,N)   \cr
&+ \ k_7(4,N) \ \Theta(3,\Lambda^+,N) \ \Theta(2,\Lambda^+,N)^2  } \eqno(A2.1)  $$

$$ \eqalign{
&k_{43}(1,N)= - {8640 \ k_{43}(5,N) \ \over (N + 1)} \ (N - 1) \ N \ (N + 2) \ (N + 3) \ (N^4 + 4 \ N^3 + 5 \ N^2 + 2 \ N + 60)  \cr
&k_{43}(2,N)={8640 \ k_{43}(5,N) \over (N + 1)^2} \ (N - 1) \ (N + 3) \ (6 \ N^6 + 36 \ N^5 + 13 \ N^4 - \cr
& \ \ \ \ \ \ \ \ \ \ \ \ \ \ \ \ \ \ \ \ \ \ \ \ \ \ \ \ \ \ \ \ \ \ \ \ \ \ \
\ \ \ \ \ \ \ \ \ \ \ \ \ \ \ \ \ \ \ \ \ \ \ \ \ \ \ \ \ \ \ \ \ \ \ \ 188 \ N^3 - N^2 + 470 \ N + 840) \cr
&k_{43}(3,N)={720 \ k_{43}(5,N) \over (N + 1)^2} \ (N - 1) \ (N + 3) \ (N^8 + 8 \ N^7 + 16 \ N^6 - 16 \ N^5 + 681 \ N^4 + \cr
& \ \ \ \ \ \ \ \ \ \ \ \ \ \ \ \ \ \ \ \ \ \ \ \ \ \ \ \ \ \ \ \ \ \ \ \ \ \ \
\ \ \ \ \ \ \ \ \ \ \ \ \ \ \ \ \ \ \ \ \ \ \ \ \ \ \ \ \ \ \ \ \ \ \ \ \ \ 2980 \ N^3 - 986 \ N^2 - 8060 \ N - 8400) \cr
&k_{43}(4,N)= - {720 \ k_{43}(5,N) \over (N + 1)} \ (N - 1) \ (N + 3) \ (2 \ N^6 + 12 \ N^5 + 121 \ N^4 + \cr
& \ \ \ \ \ \ \ \ \ \ \ \ \ \ \ \ \ \ \ \ \ \ \ \ \ \ \ \ \ \ \ \ \ \ \ \ \ \ \
 \ \ \ \ \ \ \ \ \ \ \ \ \ \ \ \ \ \ \ \ \ \ \ \ \ \ \ \ \ \ \ \ \ \ \ \ \ \ 404 \ N^3 - 957 \ N^2 - 2690 \ N + 4200)  } $$
 \ \ \ \ \ \ \ \ \ \ \ \ \ \ \ \ \
 \ \ \ \ \ \ \ \ \ \ \ \ \ \ \ \ \
$$ \eqalign{ P_{43}(\Lambda^+,N) &=
  \ k_{43}(1,N) \ \Theta(7,\Lambda^+,N)                             \cr
&+ \ k_{43}(2,N) \ \Theta(5,\Lambda^+,N) \ \Theta(2,\Lambda^+,N)     \cr
&+ \ k_{43}(3,N) \ \Theta(4,\Lambda^+,N) \ \Theta(3,\Lambda^+,N)     \cr
&+ \ k_{43}(4,N) \ \Theta(3,\Lambda^+,N) \ \Theta(2,\Lambda^+,N)^2   \cr
&+ \ k_{43}(5,N) \ \Theta(3,\Lambda^+,N)  } \eqno(A2.2)  $$

$$ \eqalign{
&k_{52}(1,N)= - {24 \ k_{52}(6,N) \over g_{52}(N)} \ (N - 3) \ (N - 2) \ (N - 1) \ N \ (N + 2) \ (N + 3) \ (N + 4) \ (N + 5) \cr
& \ \ \ \ \ \ \ \ \ \ \ \ \ \ \ \ \ \ \ \ \ \ \ \ \ \ \ \ \ \
\ \ \ \ \ \ \ \ \ \ \ \ \ \ \ \ \ \ \ \ \ \ \ \ \ \ \ \ \ \
 \ \ \ \ \ \ \ \ \ \ \ \ \ \ \ \ \ \ \ \ \ \ \ \ \ \
(N^4 + 4 \ N^3 + 17 \ N^2 + 26 \ N - 96)            \cr
&k_{52}(2,N)={12 \ k_{52}(6,N)  \over 5 \ g_{52}(N) (N + 1)} \ (N - 3) \ (N - 2) \ (N - 1) \ (N + 3) \ (N + 4) \ (N + 5) \cr
& \ \ \ \ \ \ \ \ \ \ \ \ \ \ \ \ \ \ \ \ \ \ \ \ \ \ \ \ \ \ \ \ \ \ \
(N^8 + 8 \ N^7 + 32 \ N^6 + 80 \ N^5 + 515 \ N^4 + 1676 \ N^3 + 1648 \ N^2 + 72 \ N - 10080)                                \cr
&k_{52}(3,N)={24 \ k_{52}(6,N) \over g_{52}(N) \ (N + 1)} \ (N - 3) \ (N - 2) \ (N - 1) \ (N + 3) \ (N + 4) \ (N + 5)     \cr
& \ \ \ \ \ \ \ \ \ \ \ \ \ \ \ \ \ \ \ \ \ \ \ \ \ \ \ \ \ \ \ \ \ \ \
(6 \ N^6 + 36 \ N^5 + 13 \ N^4 - 188 \ N^3 - N^2 + 470 \ N + 840)                                                           \cr
&k_{52}(4,N)= - {12 \ k_{52}(6,N) \over g_{52}(N)} \ (N - 3) \ (N - 2) \ (N - 1) \        \cr
& \ \ \ \ \ \ \ \ \ \ \ \ \ \ \ \ \ \ \ \ \ \ \ \ \ \ \ \ \ \ \ \ \ \ \
(N + 3) \ (N + 4) \ (N + 5) \ (N^6 + 6 N^5 - 6 N^4 - 64 N^3 + 281 N^2 + 706 N - 840)        \cr
&k_{52}(5,N)= - {k_{52}(6,N) \over 5 \ g_{52}(N)} \ (N - 5) \ (N - 4) \ (N - 3) (N - 2) \ (N - 1) \ N \ (N + 1)^2 \ (N + 2) \cr
& \ \ \ \ \ \ \ \ \ \ \ \ \ \ \ \ \ \ \ \ \ \ \ \ \ \ \ \ \
\  \ \ \ \ \ \ \ \ \ \ \ \ \ \ \ \ \ \ \ \ \
(N + 3) \ (N + 4) \ (N + 5) \ (6 + N) \ (N + 7) \ (N^2 + 2 \ N + 6)  } $$
\ \ \ \ \ \ \ \ \ \ \ \ \ \ \ \ \
\ \ \ \ \ \ \ \ \ \ \ \ \ \ \ \ \
$$ \eqalign{ P_{52}(\Lambda^+,N) &=
   \ k_{52}(1,N) \ \Theta(7,\Lambda^+,N)                                  \cr
&+ \ k_{52}(2,N) \ \Theta(5,\Lambda^+,N) \ \Theta(2,\Lambda^+,N)          \cr
&+ \ k_{52}(3,N) \ \Theta(4,\Lambda^+,N) \ \Theta(3,\Lambda^+,N)          \cr
&+ \ k_{52}(4,N) \ \Theta(3,\Lambda^+,N) \ \Theta(2,\Lambda^+,N)^2        \cr
&+ \ k_{52}(5,N) \ \Theta(5,\Lambda^+,N)                                  \cr
&+ \ k_{52}(6,N) \ \Theta(3,\Lambda^+,N) \ \Theta(2,\Lambda^+,N)     } \eqno(A2.3)  $$

$$ \eqalign{
&k_{322}(1,N)={34560 \ k_{322}(7,N) \over g_{322}(N)} \ (N - 1) \ N \ (N + 1) \ (N + 2) \ (N + 3) \ (2 N^2 + 4 N - 15)                        \cr
&k_{322}(2,N)= - {8640 \ k_{322}(7,N) \over g_{322}(N)} \ (N - 1) \ (N + 3) \ (N^6 + 6 \ N^5 - 6 \ N^4 -                                      \cr
& \ \ \ \ \ \ \ \ \ \ \ \ \ \ \ \ \ \ \ \ \ \ \ \ \ \ \ \ \ \ \ \ \ \ \ \ \ \ \ \ \ \ \ \ 64 \ N^3 + 281 \ N^2 + 706 \ N - 840)                 \cr
&k_{322}(3,N)= - {1440 \ k_{322}(7,N) \over g_{322}(N)} \ (N - 1) \ (N + 3) \ (2 \ N^6 + 12 \ N^5 + 121 \ N^4 +                               \cr
& \ \ \ \ \ \ \ \ \ \ \ \ \ \ \ \ \ \ \ \ \ \ \ \ \ \ \ \ \ \ \ \ \ \ \ \ \ \ \ \ \ \ \ \ \ 404 \ N^3 - 957 \ N^2 - 2690 \ N + 4200)            \cr
&k_{322}(4,N)={720 \ k_{322}(7,N) \over g_{322}(N) (N + 1)} \ (N - 1) \ (N + 3) \ (N^8 + 8 \ N^7 - 3 \ N^6 - 130 \ N^5 +                      \cr
& \ \ \ \ \ \ \ \ \ \ \ \ \ \ \ \ \ \ \ \ \ \ \ \ \ \ \ \ \ \ \ \ \ \ \ \ \ \ 109 \ N^4 + 1452 \ N^3 + 5113 \ N^2 + 6890 \ N - 4200)            \cr
&k_{322}(5,N)={720 \ k_{322}(7,N) \over g_{322}(N)} (N - 5) \ (N - 4) \ (N - 1) \ N \ \ (N + 1) \ (N + 2)                                     \cr
& \ \ \ \ \ \ \ \ \ \ \ \ \ \ \ \ \ \ \ \ \ \ \ \ \ \ \ \ \ \ \ \ \ \ \ \ \ \ \ \ \ \ \ \ \ \ (N + 3) \ (N + 6) \ (N + 7) \ (N^2 + 2 \ N - 1)   \cr
&k_{322}(6,N)= - {120 \ k_{322}(7,N) \over g_{322}(N)} (N - 5) \ (N - 4) \ (N - 1) \ N \ (N + 2) \ (N + 3) \ (N + 6)                          \cr
& \ \ \ \ \ \ \ \ \ \ \ \ \ \ \ \ \ \ \ \ \ \ \ \ \ \ \ \ \ \ \ \ \ \ \ \ \ \ \ \ \ \ \ \ \ \ \ (N + 7) \ (N^4 + 4 \ N^3 + 6 \ N^2 + 4 \ N + 25)  } $$
\ \ \ \ \ \ \ \ \ \ \ \ \ \ \ \ \
\ \ \ \ \ \ \ \ \ \ \ \ \ \ \ \ \
$$ \eqalign{
P_{322}(\Lambda^+,N) &=
   \ k_{322}(1,N) \ \Theta(7,\Lambda^+,N)                            \cr
&+ \ k_{322}(2,N) \ \Theta(5,\Lambda^+,N) \ \Theta(2,\Lambda^+,N)    \cr
&+ \ k_{322}(3,N) \ \Theta(4,\Lambda^+,N) \ \Theta(3,\Lambda^+,N)    \cr
&+ \ k_{322}(4,N) \ \Theta(3,\Lambda^+,N) \ \Theta(2,\Lambda^+,N)^2  \cr
&+ \ k_{322}(5,N) \ \Theta(5,\Lambda^+,N)                            \cr
&+ \ k_{322}(6,N) \ \Theta(3,\Lambda^+,N) \ \Theta(2,\Lambda^+,N)    \cr
&+ \ k_{322}(7,N) \ \Theta(3,\Lambda^+,N) } \eqno(A2.4) $$
where
$$ \eqalign{
&g_7(N) \equiv \prod_{i=-5}^7 (N + i) \cr
&g52(N)=(N^2 + 2 \ N - 1) \ g_7(N) \cr
&g43(N)=(N + 1) \ g_7(N)           \cr
&g322(N)=(5 \ N^2 + 10 \ N + 11) \ g_7(N) } $$

(2) The 4 eigenvalue polinomials in order 6 \ (=4+2=3+3=2+2+2)

$$ \eqalign{
&k_6(1,N)= - {30240 \ k_6(5,N) \over g_6(N)} \ (N^4 + 4 \ N^3 + 21 \ N^2 + 34 \ N + 24) \cr
&k_6(2,N)={181440 \ k_6(5,N) \over g_6(N) \ (N + 1)} \ (N - 1) \ (N + 3) \ (N^2 + 2 \ N + 6)  \cr
&k_6(3,N)={30240 \ k_6(5,N) \over g_6(N) \ (N + 1)} \ (3 \ N^4 + 12 \ N^3 + 7 \ N^2 - 10 \ N + 72) \cr
&k_6(4,N)= - {211680 \ k_6(5,N) \over g_6(N)} \ ( N^2 + 2 \ N - 6) }  $$
\ \ \ \ \ \ \ \ \ \ \ \ \ \ \ \ \
\ \ \ \ \ \ \ \ \ \ \ \ \ \ \ \ \
$$ \eqalign{ P_6(\Lambda^+,N) &=
   \ k_6(1,N) \ \Theta(6,\Lambda^+,N)                                            \cr
&+ \ k_6(2,N) \ \Theta(4,\Lambda^+,N) \ \Theta(2,\Lambda^+,N)                    \cr
&+ \ k_6(3,N) \ \Theta(3,\Lambda^+,N)^2                                          \cr
&+ \ k_6(4,N) \ \Theta(2,\Lambda^+,N)^3                                          \cr
&+ \ k_6(5,N)  } \eqno(A2.5) $$

$$ \eqalign{
&k_{33}(1,N)= - {3024 \ k_{33}(5,N) \over g_6(N)} \ (3 \ N^4 + 12 \ N^3 + 7 \ N^2 - 10 \ N + 72)  \cr
&k_{33}(2,N)={45360 \ k_{33}(5,N) \over N \ (N + 1) \ (N + 2) \ g_6(N)} \ (N^6 + 6 N^5 + 5 N^4 - 20 N^3 - 20 N^2 + 16 N + 96) \cr
&k_{33}(3,N)={1008 \ k_{33}(5,N) \over N \ (N + 1) \ (N + 2) \ g_6(N)} \ (N^8 + 8 N^7 - 112 N^5 + 127 N^4 + \cr
& \ \ \ \ \ \ \ \ \ \ \ \ \ \ \ \ \ \ \ \ \ \ \ \ \ \ \ \ \ \ \ \ \ \ \ \ \ \
 \ \ \ \ \ \ \ \ \ \ \ \ \ \ \ \ \ \ \ \ \ \ 1404 N^3 + 580 N^2 - 2032 N - 3840) \cr
&k_{33}(4,N)= - {12096 \ k_{33}(5,N) \over N \ (N + 2) \ g_6(N)} \ (4 \ N^4 + 16 \ N^3 - 35 \ N^2 - 102 \ N + 180) } $$
\ \ \ \ \ \ \ \ \ \ \ \ \ \ \ \ \
\ \ \ \ \ \ \ \ \ \ \ \ \ \ \ \ \
$$ \eqalign{ P_{33}(\Lambda^+,N) &=
   \ k_{33}(1,N) \ \Theta(6,\Lambda^+,N)                                           \cr
&+ \ k_{33}(2,N) \ \Theta(4,\Lambda^+,N) \ \Theta(2,\Lambda^+,N)                   \cr
&+ \ k_{33}(3,N) \ \Theta(3,\Lambda^+,N)^2                                         \cr
&+ \ k_{33}(4,N) \ \Theta(2,\Lambda^+,N)^3                                         \cr
&+ \ k_{33}(5,N)  } \eqno(A2.6) $$

$$ \eqalign{
&k_{42}(1,N)={483840 \ k_{42}(8,N) \over g_{42}(N)} \ (N - 1) \ (N + 3) \ (N^2 + 2 N + 6)        \cr
&k_{42}(2,N)= - {60480 \ k_{42}(8,N) \over g_{42}(N) \ N \ (N + 1) \ (N + 2)} \ (N - 1) \ (N + 3) \  \cr
& \ \ \ \ \ \ \ \ \ \ \ \ \ \ \ \ \ \ \ \ \ \ \ \ \ \ \ \ \ \ \ \ \ \ \ \ \ \ \ \ \ \ \ \ \ \ \ \ \ \ \ \
(N^2 + 2 \ N + 6) \ (120 + 2 \ N + 5 \ N^2 + 4 \ N^3 + N^4)   \cr
&k_{42}(3,N)= - {1209600 \ k_{42}(8,N) \over g_{42}(N) \ N \ (N + 1) \ (N + 2)} (N^6 + 6 \ N^5 + 5 \ N^4 - 20 \ N^3 - 20 \ N^2 + 16 \ N + 96) \cr
&k_{42}(4,N)={60480 \ k_{42}(8,N) \over g_{42}(N) \ N \ (N + 2)} (N - 1) (N + 3) (2 \ N^4 + 8 \ N^3 - 25 \ N^2 - 66 \ N + 360) \cr
&k_{42}(5,N)={5040 \ k_{42}(8,N) \over g_{42}(N)} \ (N - 4) \ (N - 3) \ (N + 1)^2 \ (N + 5) \ (N + 6) \ (N^2 + 2 \ N + 2) \cr
&k_{42}(6,N)= - {5040 \ k_{42}(8,N) \over g_{42}(N)} \ (N - 4) \ (N - 3) \ (N + 1) \ (N + 5) \ (N + 6) \ (2 \ N^2 + 4 \ N - 1)   \cr
&k_{42}(7,N)= - {84 \ k_{42}(8,N) \over g_{42}(N)} \ (N - 4) \ (N - 3) \ (N - 2) \ (N - 1) \ (N + 1)^2 \  \cr
& \ \ \ \ \ \ \ \ \ \ \ \ \ \ \ \ \ \ \ \ \ \ \ \ \ \ \ \ \ \ \ \ \ \ \ \ \ \ \ \ \ \ \ \ \ \ \ \ \ \ \ \
(N + 3) \ (N + 4) \ (N + 5) \ (N + 6)  }  $$
\ \ \ \ \ \ \ \ \ \ \ \ \ \ \ \ \
\ \ \ \ \ \ \ \ \ \ \ \ \ \ \ \ \
$$ \eqalign{ P_{42}(\Lambda^+,N) &=
   \ k_{42}(1,N) \ \Theta(6,\Lambda^+,N)                           \cr
&+ \ k_{42}(2,N) \ \Theta(4,\Lambda^+,N) \ \Theta(2,\Lambda^+,N)   \cr
&+ \ k_{42}(3,N) \ \Theta(3,\Lambda^+,N)^2                         \cr
&+ \ k_{42}(4,N) \ \Theta(2,\Lambda^+,N)^3                         \cr
&+ \ k_{42}(5,N) \ \Theta(4,\Lambda^+,N)                           \cr
&+ \ k_{42}(6,N) \ \Theta(2,\Lambda^+,N)^2                         \cr
&+ \ k_{42}(7,N) \ \Theta(2,\Lambda^+,N)                           \cr
&+ \ k_{42}(8,N)  } \eqno(A2.7) $$

$$ \eqalign{
&k_{222}(1,N)= - {483840 \ k_{222}(8,N) \over g_{222}(N)} \ (N^2 + 2 \ N - 6) \cr
&k_{222}(2,N)={51840 \ k_{222}(8,N) \over N \ (N + 1) \ (N + 2) \ g_{222}(N)} \ (N - 1) \ (N + 3) \ \cr
& \ \ \ \ \ \ \ \ \ \ \ \ \ \ \ \ \ \ \ \ \ \ \ \ \ \ \ \ \ \ \ \ \
\ \ \ \ \ \ \ \ \ \ \ \ \ \ \ \ \ \ \ \ \ \ \ \ \ \ \
(2 \ N^4 + 8 \ N^3 - 25 \ N^2 - 66 \ N + 360) \cr
&k_{222}(3,N)={276480 \ k_{222}(8,N) \over N \ (N + 1) \ (N + 2) \ g_{222}(N)} (4 \ N^4 + 16 \ N^3 - 35 N^2 - 102 \ N + 180 ) \cr
&k_{222}(4,N)= -{8640 \ k_{222}(8,N) \over N \ (N + 1)^2 \ (N + 2) \ g_{222}(N)} \ (N^8 + 8 \ N^7 - 7 \ N^6 - 154 \ N^5 - 79 \ N^4 +  \cr
& \ \ \ \ \ \ \ \ \ \ \ \ \ \ \ \ \ \ \ \ \ \ \ \ \ \ \ \ \ \ \ \ \ \ \ \ \ \ \ \ \ \ \ \ \ \
\ \ \ \ \ \ \ \ \ \ \ \ \ \ \ \ \ \ \ \ \ \ \ \ \ \ \
860 \ N^3 + 1777 \ N^2 + 1338 \ N - 3240 ) \cr
&k_{222}(5,N)= -{4320 \  k_{222}(8,N) \over g_{222}(N)} \ (N - 4) \ (N - 3) \ (N + 5) \ (N + 6) \ (2 \ N^2 + 4 \ N - 1) \cr
&k_{222}(6,N)={2160 \ k_{222}(8,N) \over (N + 1) \ g_{222}(N)} (N - 4) \ (N - 3) \ (N + 5) \ (N + 6) \ (N^4 + 4 \ N^3 - 8 \ N + 13)  \cr
&k_{222}(7,N)= - {36 \ k_{222}(8,N) \over \  g_{222}(N)} \ (N - 4) \ (N - 3) \ (N - 2) \ (N - 1) \ (N + 3) \ (N + 4) \cr
& \ \ \ \ \ \ \ \ \ \ \ \ \ \ \ \ \ \ \ \ \ \ \ \ \ \ \ \ \ \ \ \ \ \ \ \ \ \ \ \ \ \ \ \ \ \
\ \ \ \ \ \ \ \ \ \ \ \ \ \ \ \ \ \ \ \ \ \ \ \ \ \ \ \ \
(N + 5) \ (N + 6) \ (5 N^2 + 10 \ N + 11)  }   $$
\ \ \ \ \ \ \ \ \ \ \ \ \ \ \ \ \
\ \ \ \ \ \ \ \ \ \ \ \ \ \ \ \ \
$$ \eqalign{ P_{222}(\Lambda^+,N) &=
   \ k_{222}(1,N) \ \Theta(6,\Lambda^+,N)                           \cr
&+ \ k_{222}(2,N) \ \Theta(4,\Lambda^+,N) \ \Theta(2,\Lambda^+,N)   \cr
&+ \ k_{222}(3,N) \ \Theta(3,\Lambda^+,N)^2                         \cr
&+ \ k_{222}(4,N) \ \Theta(2,\Lambda^+,N)^3                         \cr
&+ \ k_{222}(5,N) \ \Theta(4,\Lambda^+,N)                           \cr
&+ \ k_{222}(6,N) \ \Theta(2,\Lambda^+,N)^2                         \cr
&+ \ k_{222}(7,N) \ \Theta(2,\Lambda^+,N)                           \cr
&+ \ k_{222}(8,N)  } \eqno(A2.8) $$
where
$$ \eqalign{
&g_6(N) \equiv \prod_{i=-4}^6 (N + i) \cr
&g_{42}(N)=(7 \ N^2 + 14 \ N + 47) \ g_6(N) \cr
&g_{222}(N)=(5 \ N^2 + 10 \ N + 23) \ g_6(N) }   $$

(3) The 2 eigenvalue polinomials in order 5 \ (=3+2)

$$ k_5(1,N)= - {k_5(2,N) \over 5 \ (N^2 + 2 N - 1) } \ (N + 1) \ (N^2 + 2 N + 6) $$

$$ \eqalign{ P_5(\Lambda^+,N) &=
   \ k_5(1,N) \ \Theta(5,\Lambda^+,N)  \cr
&+ \ k_5(2,N) \ \Theta(3,\Lambda^+,N) \ \Theta(2,\Lambda^+,N)  } \eqno(A2.9) $$
\ \ \ \ \ \ \ \ \ \ \ \ \ \ \ \ \
\ \ \ \ \ \ \ \ \ \ \ \ \ \ \ \ \
$$ \eqalign{
&k_{32}(1,N)={72 \ k_{32}(3,N) \over g_5(N) \ (N + 1)} \ (N - 1) \ N \ (N + 2) \ (N + 3) \ (N^2 + 2 \ N - 1) \cr
&k_{32}(2,N)=- {12 \ k_{32}(3,N) \over g_5(N) \ (N + 1)^2} \ (N - 1) \ N \ (N + 2) \ (N + 3) \ ( N^4 + 4 \ N^3  + 6 \ N^2 + 4 \ N + 25 ) } $$
\ \ \ \ \ \ \ \ \ \ \ \ \ \ \ \ \
\ \ \ \ \ \ \ \ \ \ \ \ \ \ \ \ \
$$ \eqalign{ P_{32}(\Lambda^+,N) &=
   \ k_{32}(1,N) \ \Theta(5,\Lambda^+,N)                           \cr
&+ \ k_{32}(2,N) \ \Theta(3,\Lambda^+,N) \ \Theta(2,\Lambda^+,N)   \cr
&+ \ k_{32}(3,N) \ \Theta(3,\Lambda^+,N) }    \eqno(A2.10)        $$
where
$$ g_5(N) \equiv \prod_{i=-3}^5 (N + i) \ \ . $$

\end